\documentclass{moriond}

\pdfoutput=1

\usepackage{amsmath,amssymb}

\def\be{\begin{equation}}
\def\ee{\end{equation}}
\def\bea{\begin{eqnarray}}
\def\eea{\end{eqnarray}}

\newcommand{\Photo}{}

\begin{document}
\vspace*{4cm}
\title{Probing Higgs-Portal Dark Matter with Weakly Interacting Mediators}

\author{S. WESTHOFF}

\address{Institut f\"ur Theoretische Physik, Universit\"at Heidelberg, 69120 Heidelberg, Germany}

\maketitle\abstracts{The hypothesis of dark matter interacting with the standard model uniquely via the Higgs portal is severely challenged by experiments. However, if dark matter is a fermion, the Higgs-portal interaction implies the presence of mediators, which can change the phenomenology significantly. This article discusses the impact of weakly-interacting mediators on the dark-matter relic abundance, direct detection, and collider searches. At the LHC, a typical signature of Higgs-portal fermion dark matter features soft leptons and missing energy, similarly to gaugino production in models with supersymmetry. We suggest to re-interpret existing gaugino searches in the context of Higgs-portal models and to extend future searches to the broader class of dark sectors with weakly-interacting fermions.}

\section{Higgs-portal fermion dark matter}
The possibility that dark matter (DM) might interact with the standard model (SM) through the Higgs portal is compelling due to its simplicity. In the case of scalar dark matter, the Higgs-portal interaction can be elementary, compatible with a dark sector that consists of the dark matter candidate only. If dark matter is a fermion $\chi$ with mass $m_{\chi}$ around the scale of electroweak symmetry breaking (EWSB), the Higgs portal is an effective interaction of mass dimension five,
\begin{equation}\label{eq:portal}
{\cal L}_{\rm eff}=\frac{\kappa_1}{\Lambda}(\bar \chi \chi) (H^\dagger H)+\frac{i \kappa_5}{\Lambda}(\bar \chi \gamma_5 \chi) (H^\dagger H).
\end{equation}
This interaction is non-renormalizable and calls for a dark sector with one or several mediator states at a scale $\Lambda$. If this new scale is situated well above the scale of EWSB, $\Lambda\gg v$ with $v=246\,\rm GeV$, the Higgs portal is the only link between the dark sector and the SM at energies $E\ll \Lambda$, with a naturally weak coupling $\kappa\,E/\Lambda\ll 1$. The phenomenology of this scenario has been investigated at colliders, direct and indirect detection experiments, and has been confronted with the observed DM relic abundance. The prevention of over-abundance sets a lower bound on the Higgs-portal interaction. The scalar coupling $\kappa_1$ induces spin-independent DM-nucleon scattering, which is strongly bounded from above by the lack of a signal at direct detection experiments. At the LHC, for $m_{\chi}<m_h/2$, the bound on the invisible Higgs decay $h\to\chi\chi$ from Higgs data sets a tight upper limit on $\kappa/\Lambda$. For $m_{\chi}>m_h/2$, however, collider signatures are suppressed by an off-shell Higgs boson. To date, this suppression prevents sensitivity to the Higgs portal above the threshold and makes future collider searches very challenging.~\cite{Craig:2014lda} A recent combined analysis of these constraints shows that viable scenarios of Higgs-portal fermion dark matter are confined to the Higgs resonance region $m_{\chi}\approx m_h/2$, where the observed relic abundance can be obtained for a small coupling $\kappa$.~\cite{Beniwal:2015sdl}

\begin{figure}
\begin{minipage}{0.25\linewidth}
\hspace*{1cm}\raisebox{1.2cm}{a)}\hspace*{-0.5cm}\centerline{\includegraphics[width=0.6\linewidth]{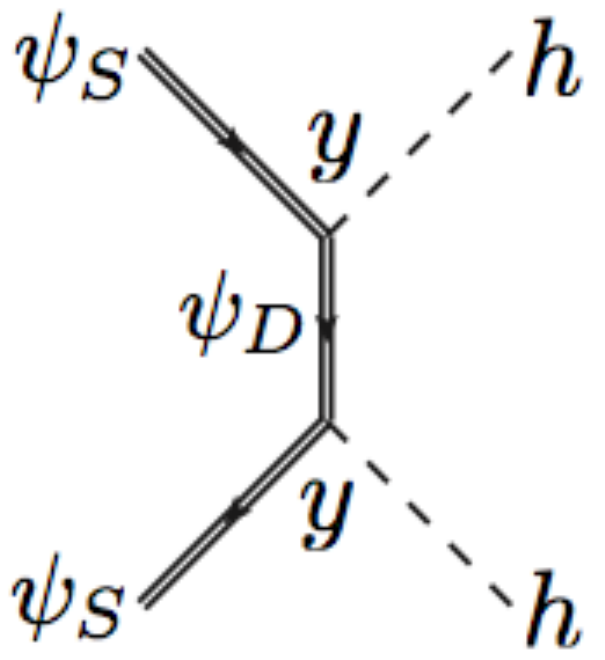}}
\end{minipage}
\hfill
\begin{minipage}{0.25\linewidth}
\raisebox{1.2cm}{b)}\hspace*{-0.5cm}\centerline{\includegraphics[width=0.6\linewidth]{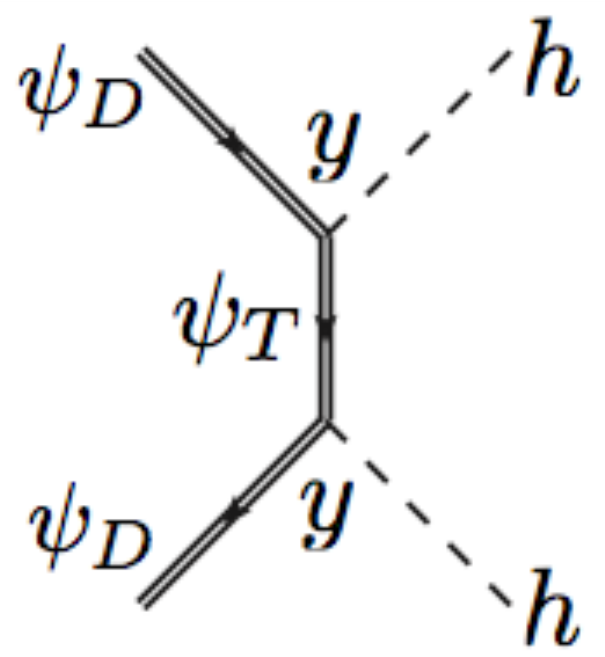}}
\end{minipage}
\hfill
\begin{minipage}{0.3\linewidth}
\hspace*{-1cm}\raisebox{0.9cm}{c)}\hspace*{0cm}\centerline{\includegraphics[width=0.85\linewidth]{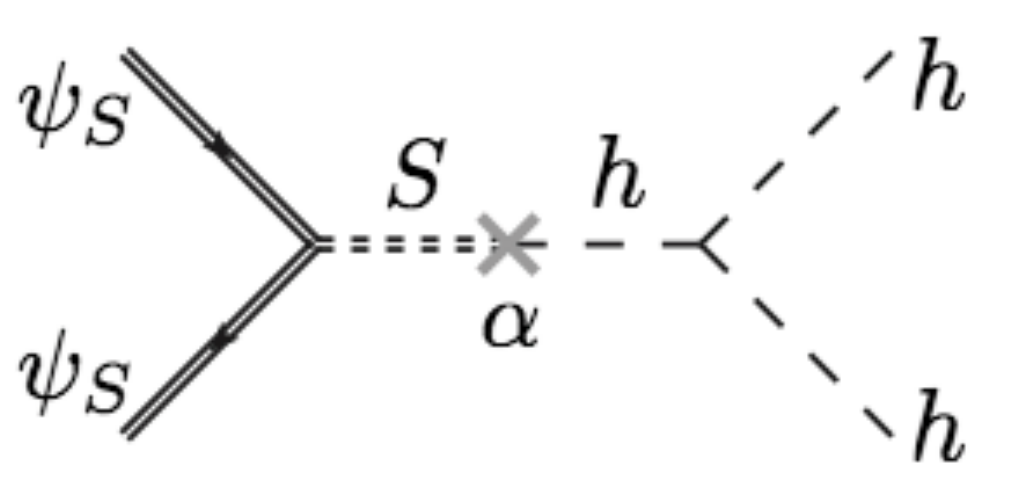}}
\end{minipage}
\caption[]{Possible realizations of the Higgs portal with fermion dark matter. a) singlet-doublet; b) doublet-triplet; c) singlet-singlet.}
\label{fig:higgs-portal-uv}
\end{figure}

If the mediator scale $\Lambda$ is around the weak scale, $\Lambda\approx v$, the Higgs portal is ``open'', i.e., the elementary couplings of the mediators become visible in DM interactions with the standard model. In particular, collider searches for Higgs-portal dark matter above the Higgs threshold become possible. The phenomenology of such a scenario depends on the nature of the mediators. We have investigated three realizations of the fermion Higgs portal,~\cite{Freitas:2015hsa} which are displayed in Fig.~\ref{fig:higgs-portal-uv}: a) a dark sector with a weak singlet fermion and a doublet fermion; b) a doublet fermion and a triplet fermion (see also Ref.~\cite{Dedes:2014hga}); c) a singlet fermion and a singlet scalar. A model with a dark triplet fermion and a quadruplet fermion has been discussed in Ref.~\cite{Tait:2016qbg}. All dark fermions have vector-like gauge interactions. In the decoupling limit $m_{\psi_D}$, $m_{\psi_T}$, $m_S\to \infty$, we recover the effective Higgs-portal interactions from Eq.~(\ref{eq:portal}).

Our goal is to study the impact of mediators on the phenomenology of the respective model, in order to determine characterstic collider signatures. A scalar mediator as in model c) mainly affects Higgs-physics observables through mixing with the Higgs boson. A fermion mediator generally introduces new interactions with the Higgs and weak gauge bosons. Here we will focus on the simplest model with fermion mediators, model a), dubbed the singlet-doublet model. The main features of this model can be extrapolated to models with fermions in larger weak multiplets. In Sec.~\ref{sec:sd-model}, we introduce the singlet-doublet model for the two cases of a Dirac or a Majorana singlet. In Sec.~\ref{sec:dd-ra}, we discuss DM-nucleon scattering and the relic abundance in the singlet-doublet model and point out the differences with the effective Higgs-portal scenario. In Sec.~\ref{sec:collider} we describe the strategy for Higgs-portal DM searches at the LHC and give an outlook on the discovery potential of future colliders.

\section{The singlet-doublet model}\label{sec:sd-model}
The dark sector of this model consists of two fermion fields, $\chi_D$ and $\chi_S$, transforming under the SM electroweak group $SU(2)_L\times U(1)_Y$ as $\chi_D\sim (2, 1/2)$ and $\chi_S\sim (1,0)$. The field $\chi_D=(\chi_D^+,\chi_D^0)$ is a doublet of Dirac fermions with vector-like gauge interactions, while the singlet $\chi_S$ can be either a Dirac or a Majorana fermion. To ensure that the lightest state in the dark sector is stable, we impose a $Z_2$ symmetry $\chi_{D,S}\to -\chi_{D,S}$, under which the SM fermions are even. For definitions and details, we refer the reader to Ref.~\cite{Freitas:2015hsa}.

\paragraph{Dirac singlet fermion.} 
If $\chi_S$ is a Dirac fermion, the particle spectrum consists of two neutral Dirac fermions, $\chi_S$ and $\chi_D^0$, and a charged Dirac fermion $\chi_D^+$. The relevant terms in the Lagrangian are 
\begin{equation}\label{eq:sd-yukawa}
{\mathcal L}_{\rm m}\supset -m_D\bar \chi_D\chi_D-m_S\bar \chi_S\chi_S - \big(y \bar \chi_D \chi_S H+\text{h.c.}\big).
\end{equation}
After EWSB, the dark Yukawa coupling $y$ introduces mixing between $\chi_S$ and $\chi_D^0$. We define the mixing angle $\theta_a$ generally as
\begin{equation}\label{eq:mixing-angle}
\sin^2\theta_a = \frac{1}{2} \biggl(1 + \frac{m_D-m_S}{\Delta m_a}\biggr),\qquad\text{with}\quad (\Delta m_a)^2 = (m_S-m_D)^2 + a(yv)^2\,.
\end{equation}
Here $a=2$, and the heavy and light mass eigenstates, $\chi_h^0$ and $\chi_l^0$, are given by
\begin{equation}\label{eq:sD:mixing}
\chi_h^0=\cos\theta_2\, \chi_S+\sin\theta_2\, \chi_D^0, \qquad \chi_l^0=-\sin\theta_2\, \chi_S+\cos\theta_2\, \chi_D^0.
\end{equation}
The corresponding mass eigenvalues are
\begin{equation}
m_{h,l}^0 = \frac{1}{2}\big(m_D+m_S \pm \Delta m_2 \big),\qquad m^+=m_D.
\end{equation}
In the basis of mass eigenstates, the interactions of the neutral fermions with the $Z$ and Higgs bosons read
\begin{align}\label{eq:dirac-zh}
\mathcal{\widehat{L}} \supset & - \frac{g}{2c_W}\Big[\cos^2\theta_2\, \bar \chi_l^{0} \gamma^\mu \chi_l^0 + \sin^2\theta_2\, \bar \chi_h^{0} \gamma^\mu \chi_h^0 + \frac{1}{2}\sin(2\theta_2)\,\big(\bar \chi_h^{0} \gamma^\mu \chi_l^0 + \bar \chi_l^{0} \gamma^\mu \chi_h^0\big)\Big]Z_\mu\\\nonumber
 & \ - \frac{y}{\sqrt{2}} \Big[ \sin(2\theta_2)\, \big(\bar \chi_h^0\chi_h^0 - \bar \chi_l^0\chi_l^0\big) + \cos(2\theta_2)\, \big(\bar \chi_h^0\chi_l^0 + \bar \chi_l^0\chi_h^0\big) \Big] h.
\end{align}
Notice that the fermion mixing affects the interactions of the DM candidate $\chi_l^0$ with the Higgs boson and induces new interactions with the $Z$ boson. The model is characterized by three parameters, which we choose to be $m_l^0$, $m_h^0-m_l^0$, and $y$.

\paragraph{Majorana singlet fermion.} \label{majsd}
If $\chi_S$ is a Majorana fermion, the dark sector consists of three Weyl fermions transforming under $SU(2)_L\times U(1)_Y$ as $\chi_S\sim (1,0)$, $\chi_D\sim (2, 1/2)$, and $\chi_D^c \sim (2, -1/2)$. In two-component notation, the relevant terms in the Lagrangian read
\begin{equation}\label{eq:Majorana:singlet}
{\cal L}_m\supset m_D\chi_{D}^c\epsilon\chi_{D} -\tfrac{1}{2}m_S\chi_S\chi_S - y (H^\dagger \chi_{D} \chi_S  - \chi_S  \chi_{D}^c \epsilon H) + \text{h.c.}.
\end{equation}
 After EWSB, the mass term for the neutral states is given by ${\cal L}_m=-\frac{1}{2}({\cal M}_{0})_{ij} \chi_i \chi_j +{\rm h.c.}$.
 In the basis $\chi_i=\{\chi_S, \chi_{D}^{c 0 }, \chi_{D}^0\}$, the mass matrix reads
\begin{equation}\label{eq:M0:doublet-singlet}
{\cal M}_0 = \begin{pmatrix}m_S & -\frac{yv}{\sqrt{2}} & \frac{yv}{\sqrt{2}} \\ 
- \frac{yv}{\sqrt{2}} & 0 & -m_D \\
 \frac{yv}{\sqrt{2}} & -m_D & 0 \end{pmatrix},
\end{equation}
which is diagonalized by the following transformation
\begin{equation}\label{eq:smd-eigensystem}
 \begin{pmatrix}\chi_h^0\\ \chi_m^0 \\
\chi_l^0 \end{pmatrix} =\begin{pmatrix}\cos\theta_4 & -\frac{1}{\sqrt{2}}\sin\theta_4 & \frac{1}{\sqrt{2}}\sin\theta_4 \\
0 & \frac{i}{\sqrt{2}} & \frac{i}{\sqrt{2}} \\
\sin\theta_4 & \frac{1}{\sqrt{2}}\cos\theta_4 & -\frac{1}{\sqrt{2}}\cos\theta_4 \end{pmatrix} \begin{pmatrix}\chi_S\\
\chi_{D}^{c0} \\ \chi_{D}^{0} \end{pmatrix}.
\end{equation}
The mixing angle $\theta_4$ is given by Eq.~(\ref{eq:mixing-angle}) with $a=4$.
The mass spectrum then reads
\begin{equation}
m_{h,l}^0=\tfrac{1}{2}\big(m_D+m_S\pm \Delta m_4 \big), \qquad m_m^0=m_D=m_+.
\end{equation}
The couplings of the neutral fermions to $Z$ and $h$ are given by 
\begin{align}\label{eq:majorana-zh}
{\cal \widehat{L}} \supset & \ i \frac{g}{2c_W} \big(\sin\theta_4\, \chi_h^{0\ast} - \cos\theta_4\, \chi_l^{0\ast}\big) \bar \sigma^\mu \chi_m^0 Z_\mu + {\rm h.c.}\\\nonumber
& - \frac{y}{2} \Big[\sin(2\theta_4)\big(\chi_h^{0}\chi_h^0 - \chi_l^{0}\chi_l^0\big) - 2 \cos(2\theta_4)\chi_h^{0}\chi_l^0 \Big]h + {\rm h.c.}.
\end{align}
Unlike in the Dirac case, the Majorana fermions have no mass-diagonal couplings to the $Z$ boson. The parameter space of this model consists of $m_l^0$, $m_m^0-m_l^0$, and $y$. This scenario corresponds to the bino-higgsino system (with decoupled wino) in the Minimal Supersymmetric Standard Model (MSSM) for $\tan\beta=1$ and $y=g'/\sqrt{2}$.

\section{Dark matter-nucleon scattering and relic abundance}\label{sec:dd-ra}
The couplings to $Z$ and Higgs bosons in Eqs.~(\ref{eq:dirac-zh}) and (\ref{eq:majorana-zh}) induce DM interactions with atomic nuclei at tree level. In the limit of zero momentum transfer the cross section for DM scattering off a nucleon $N$ is given by
\begin{equation}
 \sigma_{N}=k\,\frac{m_\chi^2 m_N^2}{(m_\chi+m_N)^2}\frac{f_N^2}{\pi},\qquad f_N = \sum_{q} G_Z^q + \sum_{q} f_{Tq}^{(N)} G_h^q \frac{m_N}{m_q} + \frac{2}{27}f_{TG}^{(N)}\sum_{Q} G_h^Q\frac{m_N}{m_Q},
\end{equation}
with $k=1(4)$ for Dirac (Majorana) dark matter, light quarks $q=u,d,s$ and heavy quarks $Q=c,b,t$, and nucleon form factors $ f_{Tq}^{(N)}$ and $f_{TG}^{(N)}$. Vector and scalar currents induce the dominant spin-independent DM interactions with the nucleus. The corresponding effective couplings for Dirac and Majorana dark matter read
\begin{align}
\text{Dirac } \chi_S + \chi_D:\  G_Z^q & =-\frac{g^2(T_q^3-2s_W^2Q_q)}{4c_W^2m_Z^2}\cos^2\theta_2, & G_h^q & =\,\frac{g}{2m_h^2}\frac{m_q}{m_W}\frac{y}{\sqrt{2}}\sin(2\theta_2);\\\nonumber
\text{Majorana } \chi_S + \chi_D:\  G_Z^q & =0, & G_h^q & =\frac{g}{4 m_h^2}\frac{m_q}{m_W}y\sin(2\theta_4).
\end{align}
For Dirac dark matter, the $Z$-mediated interaction $G_Z^q$ dominates the cross section. The non-observation of spin-independent scattering at direct detection experiments sets a strong bound on the dark-fermion mixing angle, $\cos\theta_2$. Majorana dark matter does not couple to the vector current, so that nucleon scattering is induced only by the scalar coupling $G_h^q$ from Higgs exchange. Current experiments are sensitive even to this smaller rate, yielding a strong bound on $y\sin(2\theta_4)$. In either case, to be compatible with direct detection results, the DM state must be an almost pure weak singlet, $\chi_l^0\sim \chi_S$, with a strongly suppressed dark Yukawa coupling $y$.\\

\begin{figure}
\centerline{\includegraphics[width=0.9\linewidth]{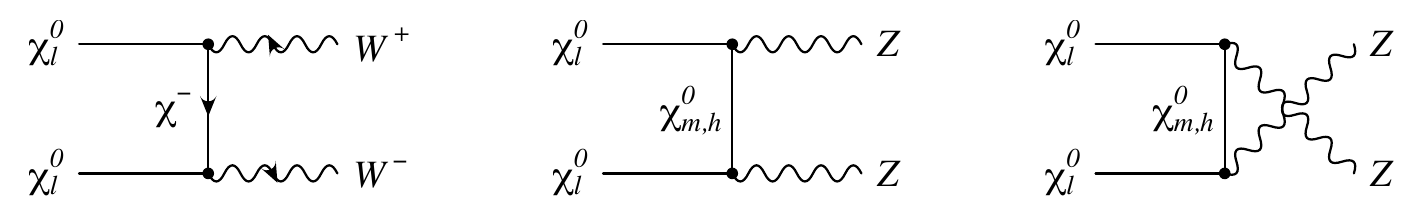}}
\caption[]{Dominant annihilation channels of Majorana fermion dark matter in the singlet-doublet model.}
\label{fig:annihilation}
\end{figure}

The suppression of electroweak couplings has important consequences on the DM relic abundance. Dirac DM annihilation proceeds dominantly through $s$-channel $Z$-boson exchange, $\chi_l^0\chi_l^0\to Z^{\ast}\to f\bar{f}$. (For $m_l^0\gtrsim 1\,\rm{TeV}$, annihilation into $WW$, $ZZ$ final states becomes relevant.) For Majorana dark matter, the dominant annihilation is into pairs of electroweak gauge bosons through $t$-channel exchange of mediators, as shown in Fig.~\ref{fig:annihilation}.
 Assuming thermal freeze-out, the strong suppression due to direct detection constraints leads to an over-abundance of dark matter, unless co-annihilation $\chi_l^0\chi^{\pm}$, $\chi_l^0\chi_m^0$ with mediator states enhances the annihilation rate. The relic abundance as observed by the Planck collaboration,~\cite{Ade:2015xua}
\begin{equation}\label{eq:relic}
\Omega_{\rm{DM}}^{\rm{Planck}} h^2 = 0.1199 \pm 0.0022,
\end{equation}
can be obtained with co-annihilation from a spectrum of dark fermions with small mass splittings $m^+ - m_l^0$, $m_{h,m}^0 - m_l^0$ of a few tens of GeV. A narrow spectrum thus results to be a typical feature of Higgs-portal dark matter with mixing dark fermions. An exception is Majorana dark matter near the Higgs resonance, $m_{\chi}\approx m_h/2$. In this case, the observed relic abundance can be obtained without co-annihilation through resonant Higgs decay into $b\bar{b}$, $W W^*$, $gg$ and $\tau^+\tau^-$ final states, and larger mass splittings are possible.\\

\begin{figure}
\begin{minipage}{0.49\linewidth}
\raisebox{0.1cm}{\centerline{\includegraphics[width=0.98\linewidth]{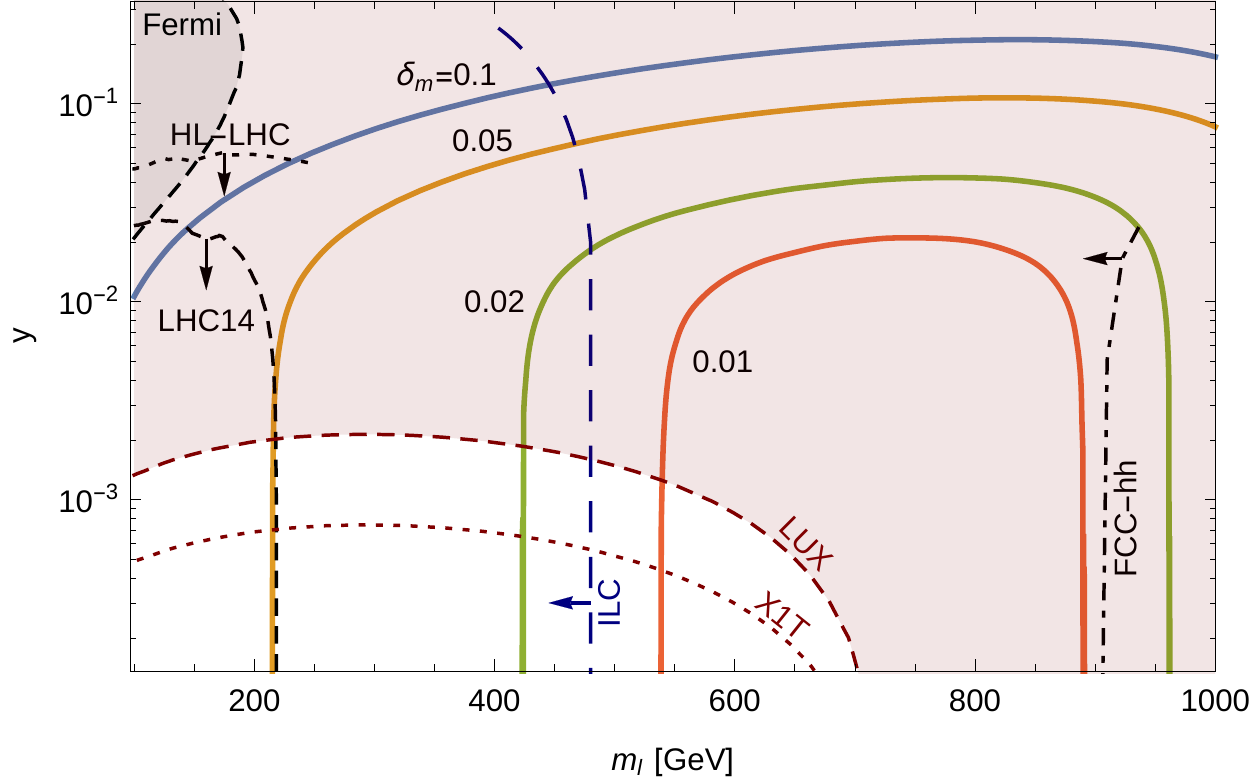}}}
\end{minipage}
\hfill
\begin{minipage}{0.49\linewidth}
\centerline{\includegraphics[width=0.98\linewidth]{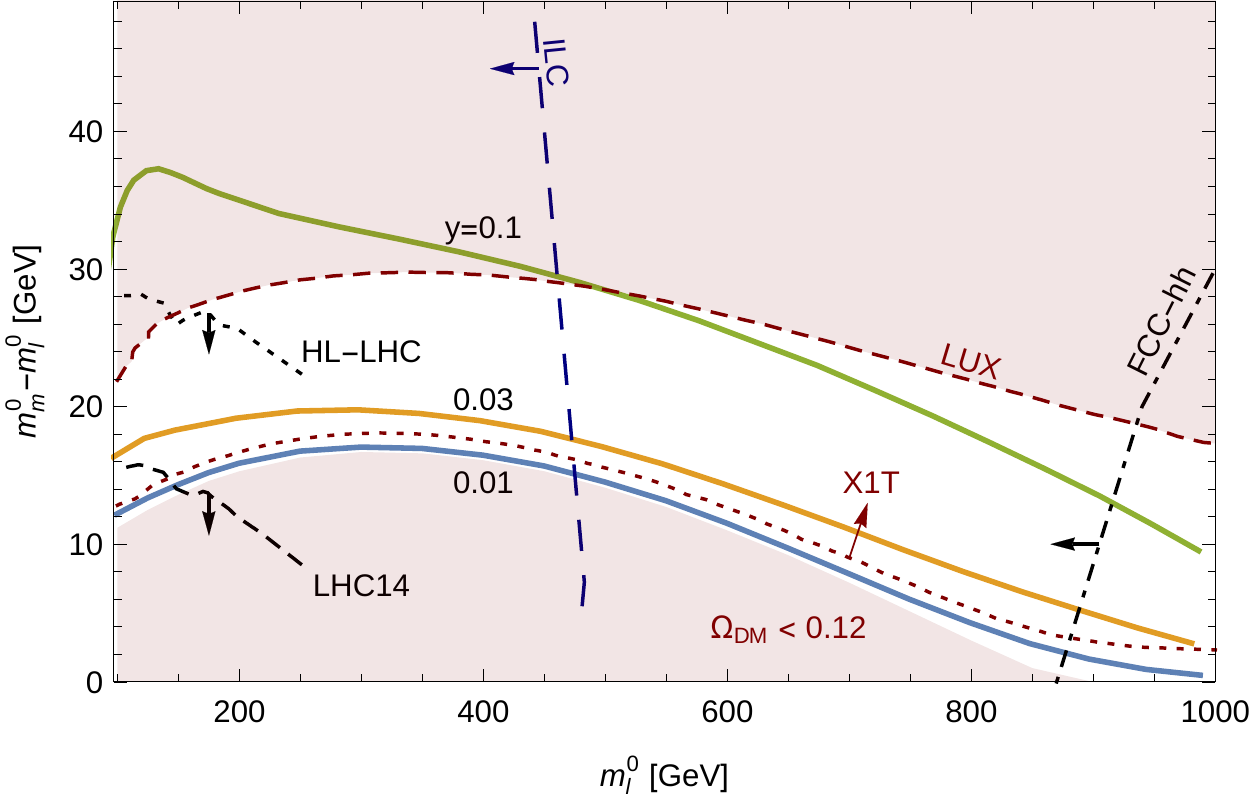}}
\end{minipage}
\caption[]{Constraints on the parameter space of the singlet-doublet model with Dirac (left) and Majorana (right) fermion dark matter. The relic density is fixed to the value obtained from the Planck mission, $\Omega_{\chi}=\Omega_{\rm{DM}}^{\rm{Planck}}$. Colored curves correspond to a constant mass splitting $\delta_m \equiv (m_h^0-m_l^0)/m_l^0$ and dark Yukawa coupling $y$, respectively. The white areas remain allowed by direct and indirect detection constraints and the relic abundance.}
\label{fig:SD-Dirac}
\end{figure}

In Fig.~\ref{fig:SD-Dirac}, we summarize the constraints from direct detection and relic abundance on the singlet-doublet model for Dirac dark matter (left) and Majorana dark matter (right). We focus on the DM mass range where co-annihilation is required to obtain the observed relic density, $100\,{\rm{GeV}} < m_l^0 < 1\,{\rm{TeV}}$. For $m_l^0 > 1\,\rm{TeV}$, the mixing between dark fermions becomes increasingly fine-tuned, given the small mass splitting. The observed relic abundance from Eq.~(\ref{eq:relic}) fixes one of the three parameters $m_l^0$, $m_{h,m}^0-m_l^0$, $y$. It also excludes the parameter range of small mass splittings (lower red area), where co-annihilation is not efficient. Upper red areas are excluded by tight bounds from LUX~\cite{Akerib:2013tjd} and other direct detection experiments. Indirect detection is not as sensitive to these scenarios, due to the suppressed DM annihilation rate. Notice, however, the exclusion of the upper-left corner of the parameter space for Dirac dark matter by recent data from the Fermi LAT.~\cite{Ackermann:2015zua} The remaining allowed parameter space (the white area) shows that Higgs-portal models with fermion dark matter around the weak scale are a viable option, if mediators are of the same scale. This is different from the decoupling scenario with heavy mediators from Eq.~(\ref{eq:portal}), where dark matter around the weak scale is excluded by direct detection and relic density observations. The main reason for this difference is the possibility of DM co-annihilation with the mediators, which changes the parametric relation between the relic density and DM-nucleon scattering.

\section{Dark fermion searches with soft leptons at colliders}\label{sec:collider}
\begin{figure}
\begin{minipage}{0.45\linewidth}
\centerline{\includegraphics[width=0.7\linewidth]{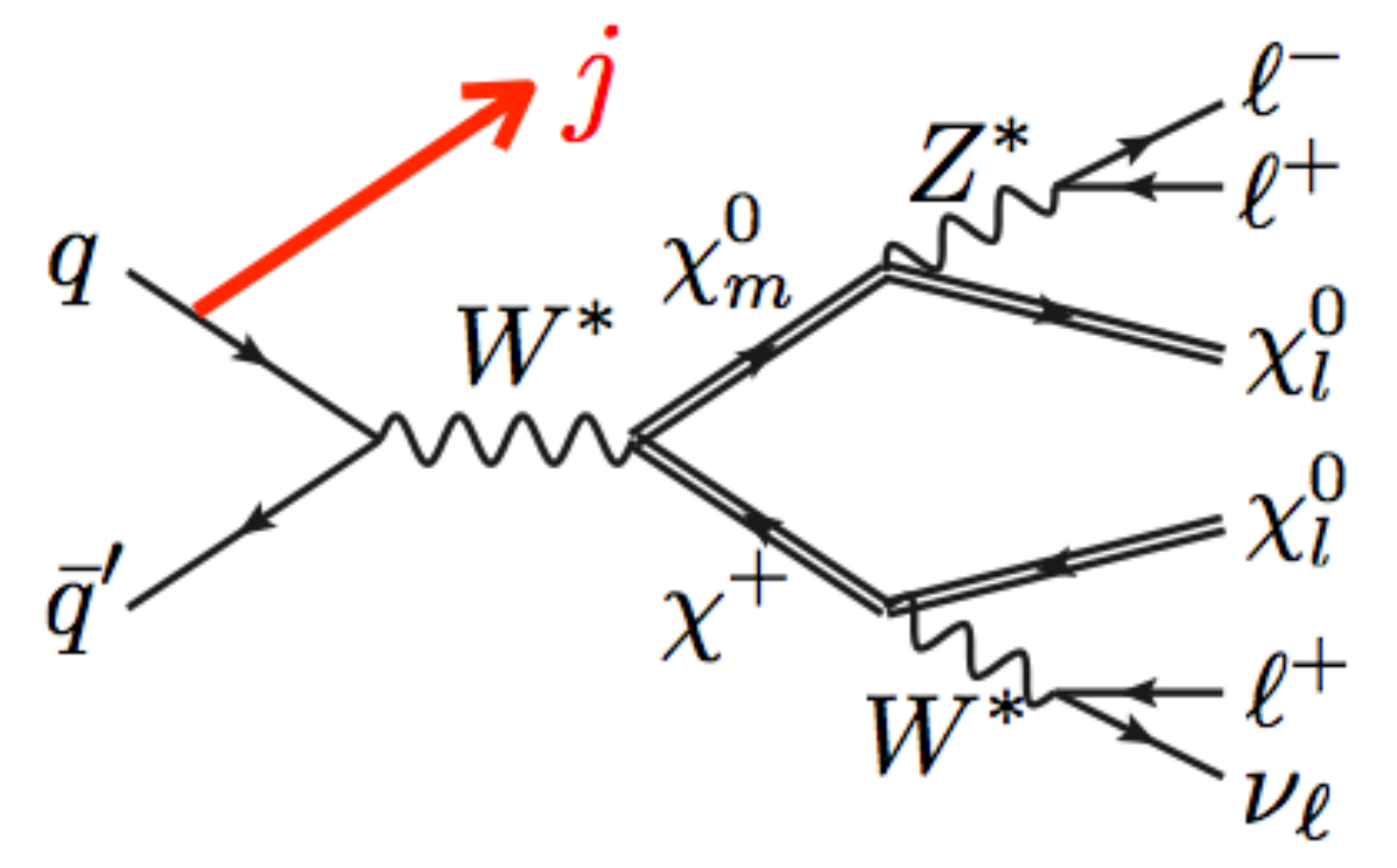}}
\end{minipage}
\hfill
\begin{minipage}{0.45\linewidth}
\centerline{\includegraphics[width=0.55\linewidth]{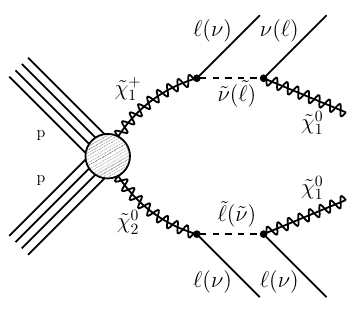}}
\end{minipage}
\caption[]{Left: typical collider signature of Higgs-portal dark matter with mixing dark fermions (here: singlet-doublet model with Majorana singlet). Right: LHC signature of supersymmetric gauginos with soft leptons.~\cite{Khachatryan:2015pot}}
\label{fig:LHC-signal}
\end{figure}

The narrow spectrum, necessary to satisfy the constraints from direct detection and relic abundance, leads to a characteristic collider signature of Higgs-portal dark matter with mixing dark fermions. Assuming DM states around the weak scale, the mediators should be resonantly produced in proton-proton collisions at LHC energies through weak interactions. The mediators then decay into DM states and leptons (or jets) through the dominant process
\begin{equation}
\begin{aligned}
q\bar{q}' &\to W^{*-} \to \chi^-\chi_{m,h}^0, \\
\chi^- &\to \chi_l^0 W^{*-} \to \chi_l^0 \ell^- \bar{\nu}_\ell, 
\quad
\chi^0_{m,h} \to \chi_l^0 Z^{*} \to \chi_l^0 \ell^- \ell^+, \qquad 
(\ell = e,\mu).
\end{aligned}
\label{eq:lhcsign}
\end{equation}
A corresponding Feynman diagram is shown in Fig.~\ref{fig:LHC-signal}, left. The small mass splittings $m^+ - m_l^0$, $m_{h,m}^0-m_l^0$ lead to soft leptons in the final state, accompanied by missing energy from DM states and neutrinos that escape the detector. Requiring an additional hard jet boosts the final state, thus helping to pass the trigger requirements of large missing energy. The signal-to-background ratio can further be enhanced by lowering the cuts on the transverse momenta of the charged leptons to be more sensitive to soft decay products.

Soft-lepton signatures have been studied and optimized for the LHC in the context of supersymmetric gauginos with compressed spectra.~\cite{Giudice:2010wb,Gori:2013ala,Schwaller:2013baa} Ref.~\cite{Schwaller:2013baa} predicts a promising signature for run II with up to three soft leptons, a hard jet and missing energy. We have re-casted their analysis for our Higgs-portal models. The results are shown in Fig.~\ref{fig:SD-Dirac} for $\sqrt{s}=14\,\rm TeV$ and luminosities of $300\,\rm fb^{-1}$ (black dashed curves) and $3000\,\rm fb^{-1}$ (black dotted curves). The LHC is expected to test the co-annihilation scenario of fermion Higgs-portal dark matter for $m_l^0\lesssim 250\,\rm GeV$.

Searches for leptons and missing energy at the LHC during run I have not lead to constraints on Higgs-portal scenarios, since the predicted lepton momenta are too soft to pass the trigger and analysis cuts. Recently, a dedicated search for supersymmetric gauginos with soft leptons has been performed with 8-TeV data.~\cite{Khachatryan:2015pot} Compressed  gauginos are one possible scenario among models with a fermionic dark sector. They lead to a similar collider phenomenology, as can be seen by comparing the final states of the processes displayed in Fig.~\ref{fig:LHC-signal}. The opportunity to test a broader class of models with soft-lepton signatures should not be missed. We therefore suggest to re-interpret current and future searches for supersymmetry with soft leptons in terms of Higgs-portal fermion dark matter. By optimizing the analysis for the parameter space of Higgs-portal models, we might be able to test the co-annihilation scenario with existing data already today.

To probe higher DM masses in our models, future colliders will be helpful. Again, we have re-casted existing projections of supersymmetry searches with soft leptons for the planned electron-positron ILC with $\sqrt{s}=1\,\rm TeV$~\cite{Berggren:2013bua} and a possible 100-TeV proton-proton collider.~\cite{Low:2014cba,Bramante:2014tba} The ILC is expected to test the DM mass range up to $m_l^0\lesssim 500\,\rm GeV$, almost half the collider energy, due to its very clean environment. A 100-TeV collider will ultimately be able to reach TeV-scale masses and thereby test these models conclusively.

\section*{Acknowledgments}
I thank the organizers of the 51st Rencontres de Moriond for inviting me to speak at this interesting and inspiring conference. Furthermore, I wish to thank Ayres~Freitas and Jure~Zupan for collaboration and sharing insight on this topic. I acknowledge funding by the Carl-Zeiss foundation through a "Junior-Stiftunsprofessur der Carl-Zeiss-Stiftung".

\end{document}